\documentclass[conference]{IEEEtran}
\IEEEoverridecommandlockouts
\usepackage{cite}
\usepackage{amsmath,amssymb,amsfonts}
\usepackage{algorithmic}
\usepackage[colorinlistoftodos]{todonotes}
\usepackage{hyperref} 
\usepackage{graphicx}
\usepackage{algorithm} 
\usepackage{booktabs}  
\usepackage{textcomp}
\usepackage{xcolor}
\def\BibTeX{{\rm B\kern-.05em{\sc i\kern-.025em b}\kern-.08em
    T\kern-.1667em\lower.7ex\hbox{E}\kern-.125emX}}
\begin{document}

\title{Mitigating Forced Oscillations in Power Systems via Data-Enabled Predictive Control \\
}

\author{
\IEEEauthorblockN{Soraya Daabak, Verena Häberle, Gabriela Hug, and Gustavo Valverde}
\IEEEauthorblockA{\textit{Power Systems Laboratory (PSL)} \\
\textit{ETH Zurich}\\
Zurich, Switzerland \\
\{sdaabak,verenhae,ghug,gustavov\}@ethz.ch}
}

\maketitle

\begin{abstract}
Sustained forced oscillations in power systems, driven by large cyclic loads such as data centers, pose a challenge to conventional power system stabilizers (PSSs), which rely on fixed tuned parameters and limited adaptability. This paper investigates the use of Data-Enabled Predictive Control (DeePC) as a data-driven alternative for damping such oscillations. DeePC constructs control actions directly from measured trajectories without requiring an explicit system model, enabling adaptation to changing operating conditions. We evaluate the performance of DeePC on a multi-machine two-area system subject to forced oscillations and compare it against a conventional PSS. The study examines the impact of different input–output configurations and the role of representative historical data in the Hankel matrix construction. Results show that DeePC can achieve superior damping. However, its effectiveness depends critically on the quality and representativeness of the underlying dataset. These findings highlight the potential of data-driven predictive control to complement or outperform conventional stabilizers in modern power systems with evolving and uncertain dynamics.
\end{abstract}


\section{Introduction}

Power system stabilizers (PSSs) are widely deployed to damp electromechanical oscillations in power systems. However, these controllers are typically not self-adaptive and are seldom retuned, as their parameter settings are often determined through offline studies and manual engineering adjustments \cite{kundur2022power,farahani2023oscillatory,musca2022power}. At the same time, emerging grid dynamics, such as sustained forced oscillations induced by large data centers running artificial intelligence (AI) workloads \cite{valverdeforcedoscillations,ko2025wide}, pose new challenges. Such oscillations can be difficult to mitigate using PSSs alone, particularly when they are poorly tuned or when operating conditions change and require extensive retuning.

Data-Enabled Predictive Control (DeePC) \cite{coulson2019deePCshallows} offers a fundamentally different, data-driven approach. By constructing predictions directly from measured trajectories organized in Hankel matrices, DeePC bypasses explicit model identification and inherently adapts to the current operating condition. This makes it particularly well suited for systems with unmodeled or time-varying dynamics, such as persistent forced oscillations whose signatures are captured in historical data.

In this paper, we investigate whether DeePC can damp forced oscillations more effectively than a conventional PSS. Specifically, we analyze the impact of different control input configurations (single versus multiple inputs) and quantify the resulting performance gains and limitations relative to a PSS. Our objectives are threefold: (i) to assess the ability of DeePC to reduce forced-oscillation amplitudes, (ii) to compare alternative output selections, and (iii) to evaluate the feasibility of operating DeePC alongside the PSS in the voltage-regulation loop of synchronous machines. The results provide insight into when and how a data-driven predictive controller can complement or outperform a PSS, while emphasizing the critical importance of informative data for closed-loop performance.

Existing studies have demonstrated the effectiveness of DeePC across a range of power system applications, including wind generator stabilization, grid synchronization, DC voltage regulation, and frequency control in wind farms \cite{markovsky2023datadriventutorial}. DeePC has also been applied to grid-connected power converters \cite{huang2019deePCconverters} and voltage source converters in HVDC systems \cite{huang2021decentralizeddataenabledpredictivecontrol}, where it has shown promising results in damping naturally occurring, weakly damped oscillations. However, these prior works primarily focus on \emph{intrinsic} system oscillations arising from poorly damped dynamics, and do not address the distinct challenge of \emph{forced oscillations} driven by persistent external disturbances. In particular, there is limited evidence on the ability of data-driven predictive control to mitigate oscillations induced by cyclic loads, such as those originating from large data centers running AI workloads \cite{valverdeforcedoscillations,ko2025wide}. These AI-driven loads introduce sustained and structured disturbances that can propagate through the grid and are often difficult to suppress using conventional control strategies. This paper addresses this gap by explicitly investigating the use of DeePC to damp forced oscillations induced by AI workloads. In doing so, we extend the scope of DeePC from stabilizing endogenous system dynamics to actively counteracting externally driven oscillatory behavior, thereby tackling an emerging and practically relevant challenge in modern power systems.

The remainder of this paper is organized as follows. Section \ref{sec:DeePC} introduces the DeePC algorithm, including the construction of the Hankel matrix from offline data and the formulation of the online optimization-based control law. Section \ref{sec:Results} presents numerical results for a multi-machine two-area system, demonstrating the superior performance of DeePC in damping forced oscillations compared to conventional PSS, along with a comparison of different input–output configurations. Finally, Section \ref{sec:Conclusions} concludes the paper.

\section{DeePC}\label{sec:DeePC}
The DeePC algorithm \cite{coulson2019deePCshallows} builds on Willems’ Fundamental Lemma \cite{willems2005note}, which establishes that any trajectory of a linear time-invariant system can be expressed as a linear combination of subsequences from a single persistently exciting input–output trajectory.

\subsection{Offline Construction of the Hankel Matrix}
Historical input $u_{\text{data}}$ and output $y_{\text{data}}$ trajectories collected during excitation are arranged in Hankel matrices $\mathcal{H}_{T_{\text{ini}}+N}(u_{\text{data}})$ and $\mathcal{H}_{T_{\text{ini}}+N}(y_{\text{data}})$ that capture the system's admissible behaviors under persistency of excitation. These matrices are partitioned into past and future segments as follows~\cite{coulson2019deePCshallows}:

\begin{equation}
\begin{pmatrix} U_p \\ Y_p \\ U_f \\ Y_f \end{pmatrix}
=
\begin{bmatrix}
\mathcal{H}_{T_{\text{ini}}}(u_{\text{data,past}}) \\
\mathcal{H}_{T_{\text{ini}}}(y_{\text{data,past}}) \\
\mathcal{H}_{N}(u_{\text{data,fut}}) \\
\mathcal{H}_{N}(y_{\text{data,fut}})
\end{bmatrix}  
\label{eq:Hankel}
\end{equation}

The dimensions of the matrices are:
\begin{align*}
U_p &\in \mathbb{R}^{T_{\text{ini}} m \times (T - L + 1)}, \quad U_f \in \mathbb{R}^{N m \times (T - L + 1)}, \\
Y_p &\in \mathbb{R}^{T_{\text{ini}} p \times (T - L + 1)}, \quad Y_f \in \mathbb{R}^{N p \times (T - L + 1)}.
\end{align*}

\noindent where $m \in \mathbb{Z}_{>0}$ is the number of inputs, $p \in \mathbb{Z}_{>0}$ is the number of outputs, $T_{\text{ini}} \in \mathbb{Z}_{>0}$ must be greater than the lag $l$ of the system, $N\in \mathbb{Z}_{>0}$ is the selected time horizon, $L=T_{\text{ini}}+N$ is the Hankel depth, and $T$ is the total number of data points in the Hankel matrix. According to Definition 4.4 of \cite{coulson2019deePCshallows}, if the Hankel matrix $\mathcal{H}_L(u)$ is of full row rank and $T\geq L$, the signal $u \in \mathbb{R}^{Tm}$ of order $L$ is persistently exciting the system. This implies that the input signal will be sufficiently rich to excite the system, yielding an output sequence that is representative of the system’s behavior.

The Hankel matrix is constructed offline and remains fixed during operation. It is computed once prior to the initial controller call and is not updated thereafter, under the assumption that the system dynamics do not change over the course of the online control application described next.

\subsection{Online Optimization-based Controller}
Rather than identifying a parametric model, DeePC infers system behavior directly from previously measured input–output data and computes optimal control actions accordingly \cite{coulson2019deePCshallows}. This formulation naturally allows the inclusion of input and output constraints to ensure safe operation. Based on the Hankel matrices constructed from past trajectories in \eqref{eq:Hankel}, DeePC solves a regularized optimization problem online at each sampling instant, where the regularization terms mitigate the impact of noise in both historical and real-time measurements \cite{mattsson2023regularization,coulson2019regularized,huang2021quadratic,dorfler2022bridging,coulson2021distributionally}, i.e., 
\begin{equation}
\begin{aligned}
\min_{g,\,u,\,y,\,\sigma_y} & 
\sum_{k=0}^{N-1} \big( \| u_k \|_R^2  + \| y_k - r_{t+k} \|_Q^2 \big )
+ \lambda_g \| g \|_2^2 + \lambda_{\sigma} \| \sigma_y \|_2^2 \\
\text{s.t.} \quad &
\begin{pmatrix}
U_p \\ Y_p \\ U_f \\ Y_f
\end{pmatrix} g =
\begin{pmatrix}
u_{\text{ini}} \\ y_{\text{ini}} + \sigma_y \\ u \\ y
\end{pmatrix}, \\
& u_k \in \mathcal{U}, \quad y_k \in \mathcal{Y}, \quad \forall k \in \{0, \dots, N-1\}.
\end{aligned}
\label{eq:regDeePC}
\end{equation}

\noindent The norm $\left\| u_k \right\|^2_R$ denotes the quadratic form $u_k^\top R u_k$, (similarly for $\|\cdot\|^2_Q$), where $R \in \mathbb{R}^{m \times m}$ is the control cost matrix, and $Q \in \mathbb{R}^{p \times p}$ is the output cost matrix that penalizes the deviations of $y_k$ from the desired reference $r_{t+k} \in \mathbb{R}^{p}$ at time $t+k$ \cite{coulson2019deePCshallows}. 

The vector $g \in \mathbb{R}^{T - L + 1}$ is a decision variable that forms a linear combination of the columns of the Hankel matrices constructed from previously recorded input--output trajectories~\cite{coulson2019deePCshallows}. This combination yields predicted future inputs $u = \begin{bmatrix} u_0^\top & \dots & u_{N-1}^\top \end{bmatrix}^\top \in \mathbb{R}^{Nm}$ and outputs $y = \begin{bmatrix} y_0^\top & \dots & y_{N-1}^\top \end{bmatrix}^\top \in \mathbb{R}^{Np}$ which are consistent with the behavioral data and the past measurements $u_{\text{ini}} \in \mathbb{R}^{T_\text{ini}m}$ and $y_{\text{ini}} \in \mathbb{R}^{T_\text{ini}p}$. In this way, DeePC implements predictive control without requiring an identified state-space model; instead, it relies directly on the recorded data.

The penalty on $g$ through parameter $\lambda_g$ is included to prevent overfitting to limited and noisy historical data \cite{coulson2019deePCshallows,coulson2019regularized}, encouraging small weights for each column of the Hankel matrix. The slack variable $\sigma_{y}$ is used to counteract the effect of measurement noise in the output, which can render the problem infeasible \cite{coulson2019deePCshallows,markovskynormsandregularization}. The parameter $\lambda_{\sigma}$ penalizes the magnitude of the slack variable in the objective function, discouraging large deviations from the measured output $y_{\text{ini}}$. 

At each time step, the optimization problem in~\eqref{eq:regDeePC} is solved using the most recent past input/output measurements $(u_{\text{ini}}, y_{\text{ini}})$, which then yields the optimal decision variable $g^\star$ and the corresponding predicted future input sequence $u^\star = U_f g^\star$. In closed-loop operation, typically only the first input $u_0^\star$ is applied to the plant. The plant then generates the next output measurement $y(t)$, which together with the applied input is appended to the past data vectors ($u_{\text{ini}},y_{\text{ini}}$), and the oldest entries are removed to update $(u_{\text{ini}}, y_{\text{ini}})$~\cite{shi2024efficientrecursivedataenabledpredictive}. By continuously updating past trajectories and optimizing, DeePC implements a receding-horizon control strategy analogous to model predictive control, while leveraging only measured data without requiring an explicit system model.  

\section{Results}\label{sec:Results}

In this section, we assess the use of DeePC in the voltage-regulation loop of a synchronous machine.  

\subsection{Simulation Setup}\label{sec:sim_setup}

\begin{figure}[t]
    \centering
    \includegraphics[width=\linewidth]{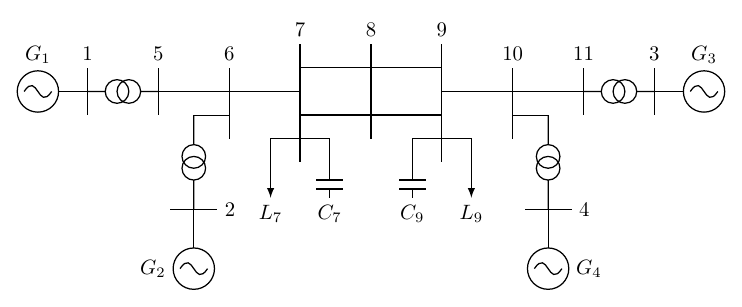}
    \caption{Multi-machine two-area test system}
    \label{fig:Kundur}
\end{figure}

DeePC is tested on the two-area system shown in Fig.~\ref{fig:Kundur} during a forced oscillation induced by a cyclic load at bus 9. The parameters of the lines, transformers, loads, synchronous machines, exciters, and Automatic Voltage Regulators (AVRs) are taken from \cite{kundur2022power}. The synchronous generators are driven by steam turbines with high-, medium-, and low-pressure sections, including a reheater. Each turbine section is modeled as a first-order system. The AVRs are equipped with a PSS represented by the classical \textit{STAB1} model, which consists of the gain $K_{PSS}$, a washout filter, and two lead-lag blocks. The PSS input is the rotor speed measurement, and the output $V_{PSS}$ goes to the summing point of the AVR. The stabilizers were tuned using the P-Vref method~\cite{krajacic2025pss}

The loads at buses 7 and 9 are voltage-dependent, totaling 2734 MW and 200 MVAr at the initial operating point \cite{kundur2022power}. The cyclic load accounts for 5\% of the active power demand at bus 9. It operates at unity power factor and has demand fluctuations of $\pm$ 50 MW around a mean power demand of 88.35 MW. The fluctuations follow a square-wave pattern at 0.55 Hz, close to the system's inter-area mode frequency. 

All time-domain simulations were run using a Python interface \cite{pyramses} to the RAMSES simulation engine~\cite{Aristidou2016}; the optimizations were solved by Gurobi~\cite{gurobi}. 

\subsection{Application of DeePC}\label{sec:appli_deepc}

\begin{figure}
    \centering
    \includegraphics[width=\linewidth]{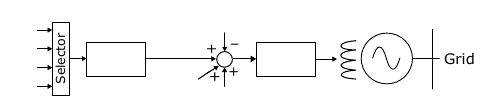}
   \put(-58,45){$G_{1}$}
    \put(-245,40){$\omega_{1}$}
    \put(-245,29){$P_{1}$}
    \put(-245,19){$Q_{1}$}
    \put(-245,8){$V_{1}$}
    \put(-140,3){$V_{PSS}$}
    \put(-140,45){$V_{1}$}
    \put(-163,9){$V_{\text{ref}}$}
    \put(-172,31){$V_{\text{DeePC}}$}
    \put(-206,23){$\text{DeePC}$}
    \put(-113,23){$\text{AVR}$}
    \caption{DeePC controller integrated into the system}
    \label{fig:DeePC_in_Syst}
\end{figure}
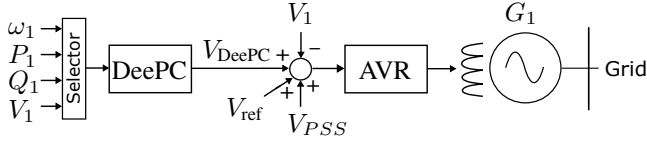

In this paper, we apply DeePC at generator $G_1$ to mitigate the forced oscillations induced by the cyclic load at bus 9. The controller receives the system outputs $y$ (e.g., rotor speed, active and reactive power output, or terminal voltage) and computes the system input $u$ (i.e., $V_{\text{DeePC}}$) added at the AVR's summing point, as depicted in Fig.~\ref{fig:DeePC_in_Syst}.  

The Hankel matrices in \eqref{eq:Hankel} were constructed by adding zero-mean white noise to the AVR's reference voltage $V_{\text{ref}}$ while tracking the system outputs. To satisfy the persistent excitation condition, we selected a standard deviation of $0.005$ pu and introduced noise samples into the system every 0.01~s. Moreover, we selected  $T=4490$, $T_{\text{ini}}=40$, and $N=100$, yielding a dataset representative of the system’s behavior. The matrices were constructed during the cyclic load operation.  

When DeePC is active, plant measurements are collected every 0.01 s to solve the optimization problem \eqref{eq:regDeePC}, with $R=10^{-4}$, $\lambda_g=0.005$, and $\lambda_c=100$. 

Due to space limitations, we only report the results for two choices of $y$: (a) the active power output $P_1$ of $G_1$ alone, and (b) $P_1$ together with the voltage magnitude $V_1$ at bus~1. However, we also tested the reactive power $Q_1$, voltage magnitude $V_1$, and the rotor speed $\omega_1$ as single outputs. The best controller performance was achieved with $P_1$ alone. Among output combinations, the best results were achieved with $P_1$ and $V_1$. 

All simulations start with the power system in steady-state conditions. The load fluctuation of 0.55 Hz starts at $t=5$ s, inducing a forced oscillation. For comparison purposes, the same Hankel matrix values were used in all scenarios; only the structure changed with the number of outputs $y$. 

DeePC minimizes the deviations of $y$ with respect to the reference values $r$, namely the output values before the cyclic load operation. 
The values of the diagonal cost matrix $Q$ were chosen as 5000 for $P_1$ and 5 for $V_1$, i.e., an output-weight ratio $Q_{P_1}/Q_{V_1}$ of 1000. 

\subsection{Simulation Results}\label{sec:appli_deepc}

In the following subsections, we analyze the system response for the two choices of 
$y$ using DeePC alone and DeePC in combination with the PSS at $G_1$. The PSSs at generators $G_2$, $G_3$, and $G_4$ remain active in all scenarios, regardless of whether the PSS at $G_1$ is enabled.  

\subsubsection{DeePC Alone}

\begin{figure}
    \centering
    \includegraphics[width=\linewidth]{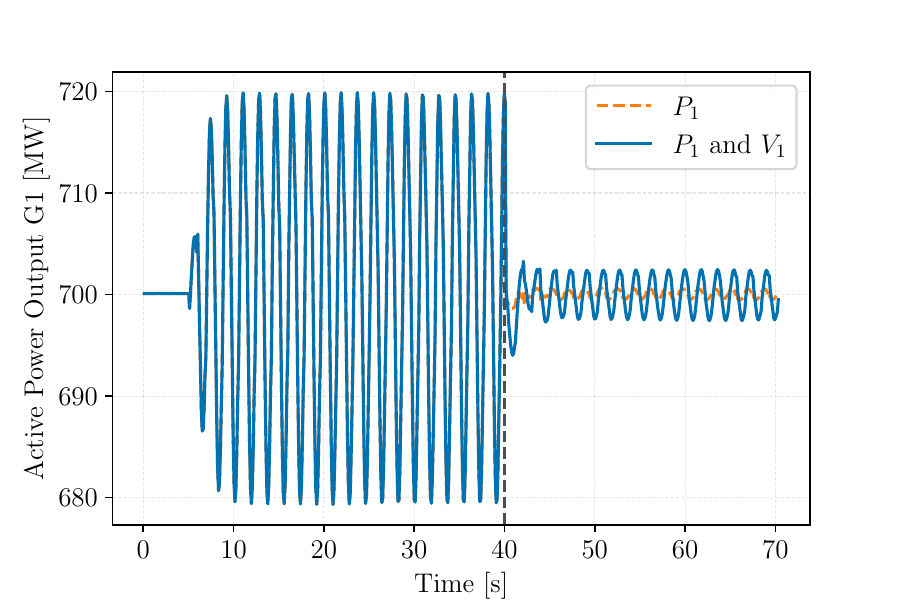}
    \caption{Active power output of generator $G_1$ before and after the DeePC (no PSS) controller.}
    \label{fig:PG1_DeePConly}
\end{figure}

In this scenario, there is no PSS at $G_1$. The damping ratio of the inter-area mode is 24.7\%, with all other modes exhibiting higher damping ratios. Fig.~\ref{fig:PG1_DeePConly} shows the forced oscillation observed at the active power output of $G_1$ before and after DeePC is activated at $t=40$~s. In the base case, without DeePC and the PSS at $G_1$, the cyclic load induces oscillations in $P_1$ with an amplitude of 20.19~MW. Once DeePC is active, the oscillations are reduced to 0.55~MW when $P_1$ is the only system output and to 2.12 MW when $P_1$ and $V_1$ are the system outputs. In the latter case, the controller simultaneously accounts for two objectives: damping oscillations in the active power output and regulating the terminal voltage. This trade-off reduces its responsiveness to oscillations in $P_1$ relative to the single-output configuration. 

\begin{figure}
    \centering
    \includegraphics[width=\linewidth]{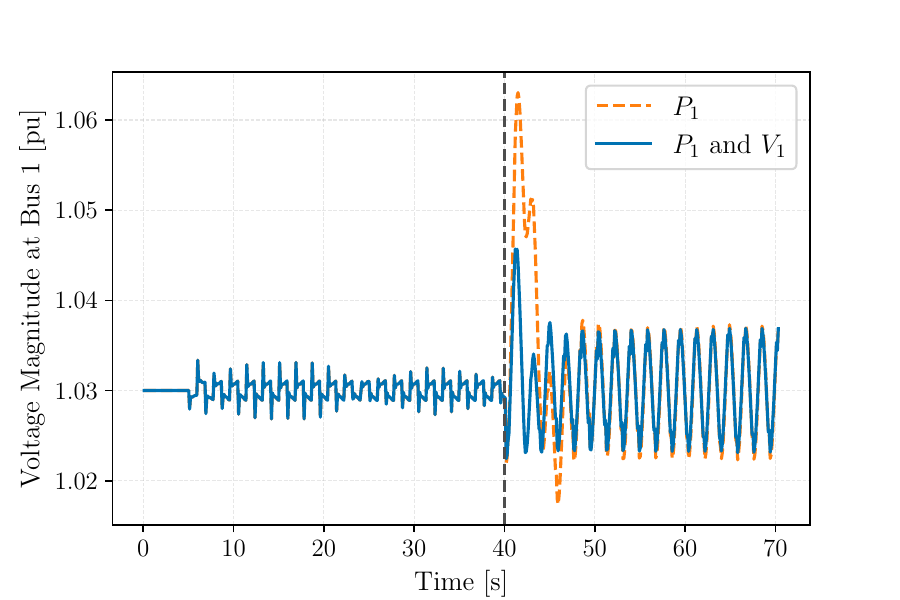}
    \caption{Voltage magnitude at Bus 1 before and after the DeePC (no PSS) controller.}
    \label{fig:V1_DeePConly}
\end{figure}

The effect on the terminal-voltage magnitude of $G_1$ is illustrated in Fig.~\ref{fig:V1_DeePConly}. Prior to DeePC activation ($t<40$~s), the oscillation amplitude is $2.6\times10^{-3}$~pu. Upon activation, a transient peak in the voltage magnitude of bus~1 appears, which is better mitigated when $V_1$ is included in the system outputs. In steady state, the oscillation amplitudes increase to $7.5\times10^{-3}$~pu and $7.0\times10^{-3}$~pu with $P_1$ alone and $P_1$ and $V_1$ as system outputs, respectively. From Figs.~\ref{fig:PG1_DeePConly} and~\ref{fig:V1_DeePConly}, it is observed that DeePC effectively modifies the machine's electrical torque to improve damping. However, this is inherently counterproductive for terminal-voltage regulation, as the DeePC signal $V_{\text{DeePC}}$ is injected into the machine's terminal-voltage control loop. 

A sensitivity analysis was conducted to examine the trade-offs between active-power oscillation mitigation and voltage regulation, by varying the output-weight ratio in the cost matrix $Q$ of \eqref{eq:regDeePC} and observing the steady-state oscillation amplitudes of $P_1$ and $V_1$. Table~\ref{tab:resultsP1V1weight_comparison} summarizes the results. Reducing the weight ratio, namely increasing the penalty on $V_1$ deviations, substantially degrades active-power oscillation mitigation while having only a minor effect on the amplitude of terminal-voltage oscillations. However, it reduces the transient peak upon DeePC activation. 

\begin{table}[ht] 
    \centering
    \caption{Steady-state oscillation amplitudes for different output-weight ratios $Q_{P_1}/Q_{V_1}$ in cost matrix $Q$ (no PSS)}
    \label{tab:resultsP1V1weight_comparison}
    \renewcommand{\arraystretch}{1.2}
    \setlength{\tabcolsep}{4pt}
    \footnotesize
    \begin{tabular}{lcccc}
        \toprule
        $Q_{P_1}/Q_{V_1}$ & 1000 & 100 & 10 & 1  \\
        \midrule
        $P_1$ [MW] & 2.12 & 6.25 & 11.46 & 13.66 \\
        $V_1$ [pu] &  0.0070  &  0.0065 &  0.0067 & 0.0067  \\
        \bottomrule
    \end{tabular}
\end{table}

\subsubsection{DeePC and PSS}

\begin{figure}
    \centering
    \includegraphics[width=\linewidth]{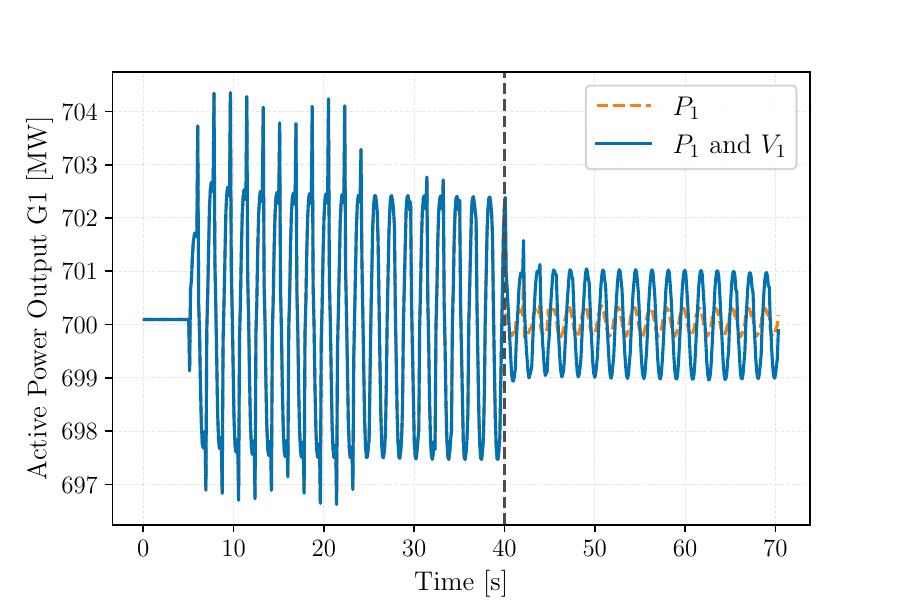}
    \caption{Active power output of generator $G_1$ before (only PSS) and after the DeePC (+ PSS) controller.}
    \label{fig:PG1_DeePCPSS}
\end{figure}

\begin{figure}
    \centering
    \includegraphics[width=\linewidth]{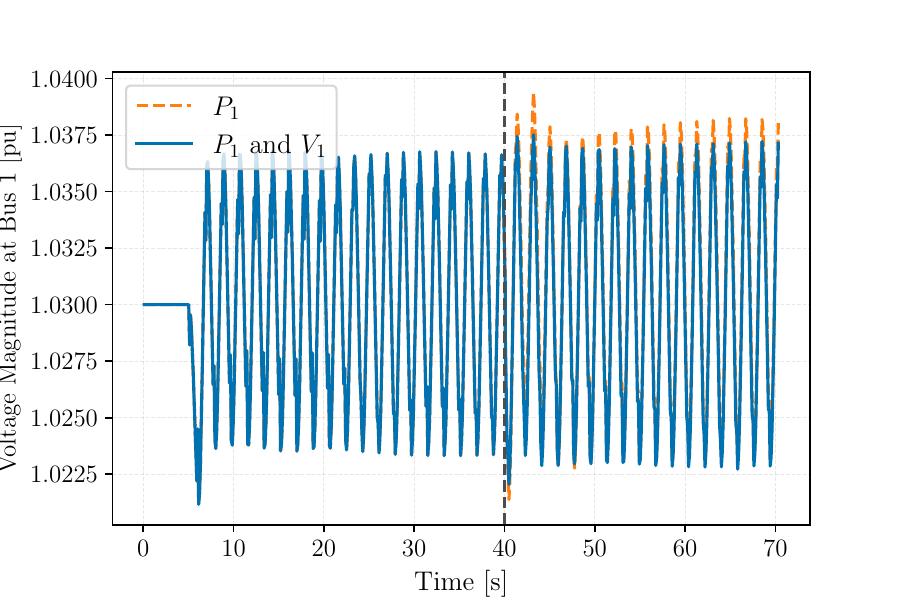}
    \caption{Voltage magnitude at Bus 1 before (only PSS) and after the DeePC (+ PSS) controller.}
    \label{fig:V1_DeePCPSS}
\end{figure}

In this scenario, there is a PSS at $G_1$ with $K_{PSS}=100$ pu/pu. The damping ratio of the inter-area mode is 68.3\%; all other modes exhibit higher damping ratios. Figs.~\ref{fig:PG1_DeePCPSS} and~\ref{fig:V1_DeePCPSS} show the active power output and terminal voltage of $G_1$, respectively. Thanks to the PSS, the initial oscillation amplitudes are smaller than in the base case. However, DeePC can further reduce the oscillation amplitude in the active power output to only 0.27 MW when $P_1$ is the only output and to 1.03 MW when $P_1$ and $V_1$ are the system outputs. Hence, DeePC, together with a well-tuned PSS, mitigates forced oscillations better compared to DeePC alone. When the PSS is present in $G_1$, DeePC has a negligible adverse effect on terminal voltage.     

\subsubsection{Impact Assessment on Other Generators}

We now investigate the impact of DeePC operating at $G_1$ on the power outputs (in MW and MVAr), rotor speeds (in pu), and terminal voltages (in pu) of all machines. Tables~\ref{tab:resultsP1} and~\ref{tab:resultsP1V1} give the steady-state oscillation amplitudes when $P_1$ and $P_1$ together with $V_1$ are used as system outputs, respectively. 
For active-power oscillation amplitudes, DeePC at $G_1$ only outperforms the PSS at $G_1$ for $P_1$ and $P_4$ damping. For the remaining machines, the PSS alone at $G_1$ is more effective than either DeePC alone or the combination of DeePC and PSS at $G_1$. For rotor-speed oscillations, DeePC alone reduces the amplitudes in $G_1$, but not in the other machines. In contrast, all voltage-magnitude oscillation amplitudes are always slightly larger when DeePC is present. Finally, regarding reactive-power oscillations, DeePC is in some cases more effective than the PSS alone.
 
\begin{table}[ht] 
    \centering
    \caption{Steady-state oscillation amplitudes with $P_1$ as system output}
    \label{tab:resultsP1}
    \renewcommand{\arraystretch}{1.2}
    \setlength{\tabcolsep}{4pt}
    \footnotesize
    \begin{tabular}{lcccc}
        \toprule
        & \textbf{Base} & \textbf{Only} & \textbf{Only} & \textbf{PSS +} \\
        \textbf{} & \textbf{Case} & \textbf{PSS} & \textbf{DeePC} & \textbf{DeePC} \\
        \midrule
        $P_1 / Q_1$ & 20.19/19.57 & 2.65/18.26 & 0.55/20.17 & 0.27/20.30 \\
        $P_2 / Q_2$ & 7.11/42.28   & 4.29/21.53  & 4.77/20.12  & 4.83/20.09  \\
        $P_3 / Q_3$ & 16.93/21.38   & 20.63/24.16  & 20.97/24.69  & 21.01/24.71  \\
        $P_4 / Q_4$ & 16.24/14.97   & 18.08/16.02  & 16.93/15.72  & 16.96/15.73  \\
        \midrule
        $\omega_1 / V_1$ & 5.3e-4/2.6e-3 & 7.7e-5/6.7e-3 & 1.7e-5/7.5e-3 & 8e-6/7.5e-3 \\
        $\omega_2 / V_2$ & 1.7e-4/5.7e-3 & 1.45e-4/7.6e-3 & 1.6e-4/8.4e-3   & 1.55e-4/8.4e-3 \\
        $\omega_3 / V_3$ & 2.4e-4/12.7e-3 & 3.0e-4/16.0e-3 & 3.0e-4/16.2e-3 & 3.0e-4/16.3e-3 \\
        $\omega_4 / V_4$ & 1.5e-4/15.8e-3 & 1.94e-4/19.6e-3 & 2.0e-4/20.1e-3 & 2.0e-4/20.2e-3 \\
        \bottomrule
    \end{tabular}
\end{table}

\begin{table}[ht] 
    \centering
    \caption{Steady-state oscillation amplitudes with $P_1$ and $V_1$ as system outputs}
    \label{tab:resultsP1V1}
    \renewcommand{\arraystretch}{1.2}
    \setlength{\tabcolsep}{4pt}
    \footnotesize
    \begin{tabular}{lcccc}
        \toprule
        & \textbf{Base} & \textbf{Only} & \textbf{Only} & \textbf{PSS +} \\
        \textbf{} & \textbf{Case} & \textbf{PSS} & \textbf{DeePC} & \textbf{DeePC} \\
        \midrule
        $P_1 / Q_1$ & 20.19/19.57 & 2.65/18.26 & 2.12/19.09 & 1.03/19.65 \\
        $P_2 / Q_2$ & 7.11/42.28   & 4.29/21.53  & 4.41/21.44  & 4.64/20.67  \\
        $P_3 / Q_3$ & 16.93/21.38   & 20.63/24.16  & 20.76/24.39  & 20.90/24.56  \\
        $P_4 / Q_4$ & 16.24/14.97   & 18.08/16.02  & 16.91/15.53  & 16.93/15.62  \\
        \midrule
        $\omega_1 / V_1$ & 5.3e-4/2.6e-3 & 7.7e-5/6.7e-3 & 5.9e-5/7.0e-3 & 3.0e-5/7.3e-3 \\
        $\omega_2 / V_2$ & 1.7e-4/5.7e-3 & 1.45e-4/7.6e-3 & 1.49e-4/7.9e-3   & 1.52e-4/8.2e-3 \\
        $\omega_3 / V_3$ & 2.4e-4/12.7e-3 & 3.0e-4/16.0e-3 & 3.0e-4/16.1e-3 & 3.0e-4/16.2e-3 \\
        $\omega_4 / V_4$ & 1.5e-4/15.8e-3 & 1.9e-4/19.6e-3 & 1.9e-4/19.8e-3 & 2.0e-4/20.0e-3 \\
        \bottomrule
    \end{tabular}
\end{table}

\section{Conclusions}\label{sec:Conclusions}

This paper investigated the use of DeePC in the AVR of a synchronous machine to mitigate the effect of forced oscillations induced by a cyclic load. We found that DeePC can effectively reduce local active-power oscillation amplitudes more effectively than a PSS. However, it increases the oscillation in the machine's terminal voltage. Including the machine's terminal voltage as a system output reduces the transient peak when DeePC is activated. However, it can interfere with the controller's ability to reduce the amplitude of active-power oscillations. The combination of DeePC and PSS yields the best results. Future work will study the feasibility of multiple agents, i.e., conventional machines or inverter-based resources, for increasing system damping with DeePC.   

\bibliographystyle{IEEEtran}
\bibliography{references}

\end{document}